\documentclass[12pt,dvips]{article}
\usepackage{amscd,verbatim}
\usepackage[all]{xy}
\usepackage{graphicx}

\usepackage{amsmath}
\usepackage[psamsfonts]{amssymb}
\usepackage{amsthm}
\usepackage{euscript}

\usepackage{latexsym}

\renewcommand{\(}{\begin{equation}}
\renewcommand{\)}{end{equation} \vspace{-.05in}\linebreak}

\newcounter{saveeqn}
\newcounter{savealpheqn}

\newcommand{\alpheqn}{\setcounter{saveeqn}{\value{equation}}%
  \stepcounter{saveeqn}\setcounter{equation}{0}%
  \renewcommand{\theequation}{\mbox{\arabic{section}.\arabic{saveeqn}
\alph{equation}}}
  \renewcommand{\)}{\end{equation}}}
\def\part#1{\frac{\partial}{\partial{#1}}}%
\def\group#1{\refstepcounter{equation}\setcounter{saveeqn}{\value{equati
on}}%
  \label{#1}\setcounter{equation}{0}%
\renewcommand{\theequation}{\mbox{\arabic{section}.\arabic{saveeqn}
\alph{equation}}}
  \renewcommand{\)}{\end{equation}}}
\newcommand{\reseteqn}{\setcounter{equation}{\value{saveeqn}}%
  \renewcommand{\theequation}{\arabic{section}.\arabic{equation}}%
  \renewcommand{\)}{\end{equation}}}

\newcommand{\aalpheqn}{\setcounter{saveeqn}{\value{equation}}%
  \stepcounter{saveeqn}\setcounter{equation}{0}%
  \renewcommand{\theequation}{\mbox{
        \Alph{subsection}.\arabic{saveeqn}\alph{equation}}}
   \renewcommand{\)}{\end{equation}}}
\newcommand{\areseteqn}{\setcounter{equation}{\value{saveeqn}}%
  \renewcommand{\theequation}{\Alph{subsection}.\arabic{equation}}%
  \renewcommand{\)}{\end{equation}}}

\renewcommand{\thefootnote}{\alph{footnote}}
\renewcommand{\(}{\begin{equation}}
\renewcommand{\)}{\end{equation}}
\newcommand{\ba}{\begin{eqnarray}}
\newcommand{\ea}{\end{eqnarray}}

\newcommand{\bp}{\mathop{\vtop{\ialign{##\crcr
   $\hfil\displaystyle{}\hfil$\crcr\noalign{\kern-13pt\nointerlineskip}
   \BIG{(}\hskip0pt\crcr\noalign{\kern3pt}}}}}
\newcommand{\cbp}{\mathop{\vtop{\ialign{##\crcr
   $\hfil\displaystyle{}\hfil$\crcr\noalign{\kern-13pt\nointerlineskip}
   \BIG{)}\hskip0pt\crcr\noalign{\kern3pt}}}}}
\newcommand{\pa}{\mathop{\vtop{\ialign{##\crcr

$\hfil\displaystyle{\oplus}\hfil$\crcr\noalign{\kern+1pt\nointerlineskip
}
   \hspace{.08in}$^{\alpha=0}$\hskip6pt\crcr\noalign{\kern3pt}}}}}

\newcommand{\R}{\ensuremath{\mathbb R}}

\newcommand{\Q}{\ensuremath{\mathbb Q}}

\newcommand{\Z}{\ensuremath{\mathbb Z}}

\def\i{\ensuremath{\dot\imath}}

\newcommand{\beq}{\begin{equation}}
\newcommand{\eeq}{\end{equation}}

\numberwithin{equation}{section}

\catcode`\@=11
\def\vereq#1#2{\lower3pt\vbox{\baselineskip1.5pt \lineskip1.5pt
\ialign{$\m@th#1\hfill##\hfil$\crcr#2\crcr\sim\crcr}}}
\catcode`\@=12

\makeatletter
\newcommand\figcaption{\def\@captype{figure}\caption}
\newcommand\tabcaption{\def\@captype{table}\caption}
\makeatother
\renewcommand{\(}{\begin{equation}}
\renewcommand{\)}{\end{equation}}

\newcommand{\bea}{\begin{eqnarray}}
\newcommand{\eea}{\end{eqnarray}}

\newcommand{\ZZ}{{\mathbb Z}}

\theoremstyle{plain}

\theoremstyle{definition}

\begin{document}

\begin{titlepage}
\begin{flushright}

\end{flushright}

\vspace{2em}
\def\thefootnote{\fnsymbol{footnote}}

\begin{center}
{\Large\bf An approach to anomalies in M-theory via KSpin}
\end{center}
\vspace{1em}

\begin{center}
\Large Hisham Sati \footnote{E-mail:
hisham.sati@yale.edu}
\end{center}

\begin{center}
\vspace{1em}
{\em { Department of Mathematics\\
Yale University\\
New Haven, CT 06520\\
USA}}\\
\end{center}

\vspace{0.5cm}
\begin{abstract}
\noindent The M-theory fieldstrength and its dual, given by the integral
lift of the left hand side of the equation of motion,  both satisfy
certain cohomological properties. We study the combined fields and
observe that the multiplicative structure on the product of the
corresponding degree four and degree eight cohomology fits into that
given by Spin K-theory. This explains some earlier results and leads
naturally to the use of Spin characteristic classes. We reinterpret
the one-loop term in terms of such classes and we
show that it is a homotopy invariant.  We argue that the 
various anomalies have natural interpretations within Spin K-theory.
In the process, mod 3 reductions play a special role. 
\end{abstract}

\vfill

\end{titlepage}
\setcounter{footnote}{0}
\renewcommand{\thefootnote}{\arabic{footnote}}

\pagebreak
\renewcommand{\thepage}{\arabic{page}}

\section{Introduction}
The non-gravitational fields in M-theory and string theory play a
major role in characterizing the topology and the global aspects of
these theories. Such fields take continuous real or complex values
in the classical supergravity limit and get quantized, so that a
priori they take values in $\Z$, in the quantum regime. The fields
take values in cohomology of the space $X$, and so classically are
in $H^*(X, \R)$ and quantum-mechanically in $H^*(X, \Z)$. An
important difference between the two cases is the presence of
torsion in the latter case and that does not exist in the former. It
is in fact this feature that gives the subtle distinction between
(generalized) cohomology theories.

\vspace{3mm} Both $G_4$ and its `dual'-- let us call it $G_8$ for
now-- involve shifts in the Pontrjagin classes. The M-theory degree
four field $G_4$ defined on an eleven-dimensional space $Y^{11}$ is
not an integral class but satisfies the shifted integrality
condition \cite{Flux} $G_4 - p_1/4 \in H^4(Y^{11}, \Z)$, where $p_1$
is the first Pontrjagin class of the tangent bundle $TY^{11}$. This
is written as \cite{Flux} \( G_4 - \lambda/2 \in H^4(Y^{11}, \Z),
\label{g4} \) where $\lambda$ is equal to half the Pontrjagin class
of the eleven-dimensional space $Y^{11}$. In comparing to
ten-dimensional string theory, described by K-theory, at the level
of partition functions, torsion fields play a major role \cite{DMW}.
In particular they lead to an anomaly for the partition function.
This is canceled in \cite{KS1} by declaring spacetime to be oriented
with respect to generalized cohomology theories beyond K-theory.

\vspace{3mm} In addition to this field that appears in the
eleven-dimensional supergravity multiplet, there is also the dual
field whose class is considered in \cite{DFM, S1, S2, S3} and has a
quantization condition of its own. This is the class given by the
integral lift of the right hand side of the equation of motion for
$G_4$ \cite{DFM}. $G_8$ is built out of a quadratic term in $G_4$
plus the one loop term, which is a polynomial expression in the
Pontrjagin classes $p_1$ and $p_2$. In \cite{S2, S3}, a distinction
is made between two fields that can be dual to $G_4$: the actual
Hodge dual $*G_4$ and the class $\Theta$ defined in \cite{DFM}.

\vspace{3mm} In this note, we investigate the multiplicative
structure on the product of the cohomology of degrees four and
eight. In particular we will show that the quadratic refinement
defined in \cite{DFM} is encoded in the multiplicative structure in
the K-theory for Spin bundles. This will motivate us to propose that
the Spin characteristic classes are the natural setting for the
above shifts. This gives an insight into the relation between $G_4$
and its `dual'. We then make connection to the classes proposed in
\cite{S1}. The calculation of the path integral involves
exponentiating the action time $2\pi i$. The requirement that the
partition function is well-defined imposes integrality properties on
the topological terms of the action. One such term is the one-loop
term (equation (\ref{I8})), whose integrality was established in
\cite{Flux} using congruence from index theory. This term takes an
interesting form when written in terms of the Spin characteristic
classes. In fact, it turns out to be essentially given by the second
Spin class, up to an interesting factor of 24 which reminds us of
other occurrences of such a factor. As a warm up to discussing the
mod $p$ reduction of the fields, we show that the one-loop term is a
homotopy invariant. The two facts strongly suggest that this term
should have a deep homotopy-theoretic meaning.

\vspace{3mm} The observation that the quadratic refinement is given
by the natural multiplication on the image of the Chern character
motivates us to seek more connections with KSpin. To make such
connections we study the mod three reductions of the fields.  The
anomalies in M-theory and type IIA string theory are encoded as
conditions on the natural bundles and the aim here is to argue for a
unified approach. We provide evidence for this from the 
quadratic structure as well as from the form of the anomalies themselves. 
This however, leaves many interesting and subtle questions open, 
such as accounting for the precise denominator factors, most 
importantly the factors $\frac{1}{2}$ and $\frac{1}{24}$. 
Nevertheless, one observation is the connection between 
$p=3$ and M-theory and between $p=2$ and string theory, 
which provides more systematic evidence for observations in our 
previous work \cite{S3}. Another theme
is the mod 24 quantization.  What we see is that this approach seems 
to treat in a unified way the anomalies in the membrane theory, in
type IIA string theory, in the fivebrane theory, and in M-theory. In terms of
classes, roughly, the M2-brane corresponds to the first Spin class and the M5-brane
\cite{5, HS} corresponds to the second Spin class. 

\vspace{3mm} 
Anomalies generally 
involve Spin bundles and so it is only natural to study them within K-theory
of such bundles.
How is Spin K-theory related to more well-known
K-theories? Given a topological space $X$, let ${\widetilde{KO}}(X)$
be the reduced $KO$ group for $X$ and let \( W~:~
{\widetilde{KO}}(X)\longrightarrow H^1(X;\ZZ_2) \times H^2(X;\ZZ_2)
\label{KO} \) be the map $W(\xi)=(w_1(\xi),w_2(\xi))$, where
$w_i(\xi)$ denotes the $i$-th Stiefel-Whitney class of $\xi \in
{\widetilde{KO}}(X)$. There is a group structure on $H^1(X;\ZZ_2)
\times H^2(X;\ZZ_2)$ making $W$ a homomorphism, i.e. a map that
preserves the group structure. Starting with a real unoriented
bundle $\xi$, the condition $w_1(\xi)=0$ turns $\xi$ into an
oriented bundle, and the condition $w_2(\xi)=0$ further makes $\xi$
a Spin bundle. Obviously then, a real $O$-bundle becomes a Spin
bundle when $W=0$, and so the kernel of $W$ is the reduced group
(see section (\ref{KSp})) $\widetilde{K{\rm Spin}}(X)$. Thus $W$
fits into the exact sequence \cite{Li}
\( 0 \longrightarrow
{\widetilde{K{\rm Spin}}}(X)=\ker W \longrightarrow
{\widetilde{KO}}(X) {\buildrel{W} \over {\longrightarrow}}
H^1(X;\ZZ_2) \times H^2(X;\ZZ_2). \)
We do not consider specific examples since $K{\rm Spin}$ of many classes 
of interesting spaces are already tabulated in \cite{Li}.

\vspace{3mm} We say that $x \in H^*(X; \Z)$ is an element of order
$r$ ($r=2, 3, 4, \cdots$) if and only if $x \neq 0$ and $r$ is the
least positive integer such that $rx\neq 0$ (if it exists). The
reduction mod $k$ induces the mapping $\rho_k~:~H^*(X;\Z) \to
H^*(X;\Z_k)$. For more background on cohomology operations, see e.g.
\cite{SE}.

\section{The One-Loop Term via Spin Characteristic Classes}
Recall that characteristic classes on a space $X$ are obtained by
pulling back to the space $X$ the universal classes from the
cohomology ring of the corresponding universal space. For oriented
vector bundles, the relevant group is $SO$ with classifying space
$BSO$. Rationally, the cohomology ring $H^*(BSO;\Q)$ is a polynomial
ring over $\Q$ generated by the universal Pontrjagin classes
$p_i\in H^{4i}(BSO;\Q)$.

\vspace{3mm} As is the case for any G-bundle, Spin bundles have a
classifying space, which is $B{\rm Spin}$, and the corresponding
characteristic classes are obtained by pulling back from that space.
More precisely, the Spin characteristic classes can be defined for
the stable class of a Spin bundle $\xi$ over a topological space
$X$, in our case an eight-, eleven- or twelve-dimensional space, by
$Q_i(\xi)=\iota^*Q_i \in H^4(X;\ZZ)$, where $\iota : X
\longrightarrow B{\rm Spin}$ is the classifying map, in the stable
range, for the bundle $\xi$. The corresponding $Q_i$ are cohomology
classes $Q_i\in H^{4i}(B{\rm Spin}; \Z)$, for $i=1,2,\cdots$.

\vspace{3mm} The Spin cohomology ring with coefficients in $\Z_2$ is
generated by the mod 2 Stiefel-Whitney classes of certain degrees
\cite{extra}. What we are interested in is integral coefficients, in
which case \( H^*(B{\rm Spin}; \Z)=\Z[Q_1,Q_2,\cdots]\oplus \gamma,
\) with $\gamma$ a 2-torsion factor, i.e. $2\gamma=0$ \cite{Thomas}.
The two degrees relevant to our discussion are \bea H^4(B{\rm Spin};
\Z)&\cong&\Z ~~~~~~~~~~~{\rm with~~generator}~~~ Q_1
\nonumber\\
H^8(B{\rm Spin}; \Z)&\cong&\Z\oplus \Z ~~~~~{\rm with~~generators}~~~ Q_1^2,
Q_2,
\eea
where $Q_1$ and $Q_2$ are determined by their relation to the Pontrjagin classes
\begin{eqnarray}
p_1&=&2Q_1 \nonumber\\
p_2&=&Q_1^2 + 2Q_2.
\label{Qs}
\end{eqnarray}
Obviously, when inverting is possible, the generators are given by
$Q_1=p_1/2$ and $Q_2=\frac{1}{2}p_2 - \frac{1}{2}(p_1/2)^2$.

\vspace{3mm} We now make the first use of the Spin classes. In
particular we use them to write the one-loop polynomial $I_8$ in a
suggestive way, and we then make connection to the classes proposed
in \cite{S1}. The one-loop polynomial of some tangent bundle $T$ is
given in terms of the Pontrjagin classes \cite{DLM}\(
I_8=\frac{p_2(T) - (p_1(T)/2)^2}{48}, \label{I8} \) where $p_1/2$ is
usually denoted $\lambda$, and represents the string class. In an
earlier work \cite{S1} we observed that $I_8$ can be written in a
way that suggests its interpretation as a Chern character
\footnote{We will come back to the character interpretation in
section \ref{KSp}.} upon using the class $\lambda$ -- which we
called $\lambda_1$ in \cite{S1}-- and another class, which we
defined as $\lambda_2 =p_2/2$, were used. This led to the expression
\( I_8=\frac{\lambda_2 - \frac{1}{2}\lambda_1^2}{24}. \label{lamb}
\)

\vspace{3mm}
 Now we proceed to write $I_8$ in terms of the Spin classes $Q_1$ and
$Q_2$ and compare the result with (\ref{lamb}). For that we simply
substitute (\ref{Qs}) to get \( I_8=\frac{Q_2}{24}. \label{Q24} \)
First, note that this expression is written entirely in terms of the
second Spin characteristic class $Q_2$ as the first one, $Q_1$,
canceled out. The relation to the classes in \cite{S1} is now
obvious. The class $\lambda_1$ is exactly $Q_1$, whose values is
half the first Pontrjagin class. The degree eight class $\lambda_2$
is then equal to $Q_2$ once $Q_1$ vanishes. This has a nice
interpretation. Since we are viewing the classes $Q_i$ as
obstructions, then it makes sense to be able to talk about the
second obstruction only after the first obstruction is absent. This
then gives the desired structure to the observations and proposal in
\cite{S1, S2} on the Spin part of the polynomials.

\section{Topological and Homotopy Invariance }
In this section we investigate whether the classes used in \cite{S1}
and the one-loop term (\ref{I8}) are topological invariant and/or
homotopy invariant. Homotopy invariance means dependence only on the
homotopy type of the manifold, and independence of the
differentiable structure. Topological invariance, on the other hand,
is the requirement of independence on the {\it choice} of a
differentiable structure. In both case, the statements depend on the
coefficient ring over which the Pontrjagin classes are taken.

\subsection{Homotopy invariance of Pontrjagin classes}
The homotopy invariance of the rational Pontrjagin classes $p_k$ depends on whether
one is considering stable or unstable bundles. For stable universal vector bundles,
$p_k \in H^{4k}(BO, \Q)$ are not homotopy invariant for $k\geq 1$, but for nonstable
vector bundles $p_k \in H^{4k}(BO[2k], \Q)$ are homotopy invariant \cite{Singh91}.
The situation for the integral Pontrjagin classes modulo
\footnote{We use $q$ instead of $p$ to denote a prime, so as not to confuse with
the several variations on $p$ used in this note.}
$q$ is as follows. For $q=2$,
$p_k$ mod 2 $=w_{2k}^2$, and since the Pontrjagin classes are homotopy invariant,
this implies that $p_k$ mod 2 are homotopy invariant. We deduce from this
that the classes $p_i/2$ used in \cite{S1} are homotopy invariant.
The integral Pontrjagin classes $p_k$ modulo $q$, where $q$ is an odd prime,
are homotopy invariant only if $q=3$. A classic result of Wu that $p_k$ mod 3
are the Wu classes $U_3^k$, which are defined in terms of the Steenrod reduced
powers (see section \ref{multwu}) implies that they are homotopy invariant.
Thus integral $p_k$ mod $q$ are not
homotopy invariant for any other $q\neq 3$ \cite{Singh96}.

\subsection{Topological invariance of Pontrjagin classes}
For a topological manifold $M$ ( for us, $Z^{12}$, $Y^{11}$,
$X^{10}$, or $M^8$), let $\Sigma_1$ and $\Sigma_2$ be two different
smooth structures and let $TM_{\Sigma_1}$ and $TM_{\Sigma_2}$ be the
corresponding tangent bundles. Associate the $k$-th Pontrjagin
classes $p_k(TM_{\Sigma_1})$ and $p_k(TM_{\Sigma_2})$ in $H^{4k}(M,
\Lambda)$. The question is whether or not $p_k(TM_{\Sigma_1})
=p_k(TM_{\Sigma_2})$. It turns out that the answer depends on the
coefficient ring $\Lambda$. For $\Lambda=\Q$, it is a classic result
of Novikov that the rational Pontrjagin classes are topological
invariants. However, this is {\it not} the case for the integral
case $\Lambda=\Z$. What about $\Lambda=\Z_q$, the ring of integers
$q$, where $q$ is any prime? In this case, as mentioned above, $p_k$
mod 3 are the Wu classes $U^k_3$, which are defined in terms of the
Steenrod reduced powers (see section \ref{multwu}) and hence are
topological invariants. This has been extended to $q=5$ in
\cite{Singh91}. Thus, for every $k\geq1$, $p_k$ mod $q$ are
topological invariant for $q=3$ and 5. However, this breaks down at
$q=7$ as then $p_2$ mod 7 is not topological invariant \cite{SS95}.

\vspace{3mm}
Since the integral Pontrjagin classes are not topological invariant, one can ask: what are the multiples
of the integral $p_k$'s that are topological invariant? The smallest possible integer $n_k$ such that
$n_k p_k$ is a topological invariant is given by $n_1=1$ and $n_2=7$ \cite{Shar}.

\subsection{Consequences for the one-loop term}
 We would like to investigate the invariance of the one-loop term (\ref{I8}) in
 the context of
the above discussion. The one-loop term is an example of a Ponrjagin
number, i.e. a polynomial of a given degree in the Pontrjagin
classes. It is known that at the rational level, the only rational
linear combination in the Ponrjagin classes that is homotopy
invariant is, up to a rational linear multiple, the Hirzebruch
L-polynomial \cite{Ka} that appears in the signature theorem.
However, the one-loop term is not quite equal to $L_2$ (see
(\ref{L2}) for the corresponding expression) and thus the polynomial
(\ref{I8}) cannot be homotopy invariant at the rational level. Thus
we are forced to study the expression modulo primes.

\vspace{3mm} In additon to homotopy invariance of the Pontrjagin
classes mod 3, there is an additional result \cite{Mad} that $p_k$
mod $2^3$ are also homotopy invariant. Thus $p_k$ mod 24 are
homotopy invariant. In particular this means that $p_2$ mod 24 is
homotopy invariant. We are still short by a factor of 2 to get the
first term in (\ref{I8}). Let us look at the analogous situation for
$p_1$. In that case, the fact that $p_1(\xi) \equiv w_2(\xi)^2$ mod
2 implied the fact that $p_1$ is even when the bundle $\xi$ is Spin,
because then $w_2(\xi)=0$. Combining the two results one has that
the first Pontrjagin class of a Spin bundle is a homotopy invariant
mod 24. Now let us see what can be said about $p_2$. Here note that
$p_2(\xi) \equiv w_4(\xi)^2$ mod 2, so that we do get the evenness
of $p_2$ provided that we have the condition $w_4(\xi)=0$, the
higher degree analog that replaces the spin condition. Note that
this is the obstruction to orientation with respect to the real
version $EO\langle 2 \rangle$ of Landweber elliptic cohomology with
two generators which appears in the study of the partition functions
\cite{KS1,KS3, S4}. Given this condition, we are then able to define
$p_2/2$ as in \cite{S1}. Going back to the one-loop term, we have so
far that the first term in (\ref{I8}) is homotopy invariant.

\vspace{3mm} What about the second term in (\ref{I8})? We consider
$p_1^2$. Since $p_k$ mod 3 are homotopy invariant then so is $p_k^m$
mod 3. In particular, then, $p_1^2$ mod 3 are homotopy invariant.
For Spin bundles $p_1$ is even so then $(\frac{1}{2}p_1)^2$ mod 3 is
a homotopy invariant and so $p_1^2$ mod 12 is a homotopy invariant.
On the other hand, from \cite{Mad}, $p_1$ mod $2^3$ is a homotopy
invariant. Combining the two results implies that $p_1^2$ mod 96 is
a homotopy invariant. Therefore, the one-loop term is a homotopy
invariant. In fact, as we have just seen, we have more: each of the
two terms separately is homotopy invariant.


\section{The Multipicative Structure on the Cohomology Ring}
\label{multwu}
In addition to the usual cohomology ring $H^*(X;\Z_q)$ which is the
direct sum of the elements in the individual degrees depending on
grading, one can also form the direct product $H^{**}(X;\Z_q)$
of the cohomology groups $H^i(X;\Z_q)$ for $i=0,1,2,\cdots$ . In this
way, the direct sum $H^{*}(X;\Z_q)$ can be thought of as being included
inside $H^{**}(X;\Z_q)$. The ring structure on both $H^{*}$ and $H^{**}$
is given by the cup-product operation. Inside the ring $H^{**}(X;\Z_q)$
one can also talk about inverting elements $x$, which is possible when
the zeroth component is nonzero in $H^0(X;\Z_q)$.

\vspace{3mm} One can form the total Steenrod reduced power operation
$P=P^0 +P^1+P^2+\cdots$ which acts as an automorphism of rings
$H^{**}(X;\Z_q) \longrightarrow H^{**}(X;\Z_q)$, and is the identity
on $H^{*}(X;\Z_q)$. The cohomology ring $H^{**}(X)$ is graded and
decomposes as $H^{**}(X)=H^{\rm even}(X) + H^{\rm odd}(X)$ where
$H^{\rm even}(X)=\prod_{m=0}^{\infty} H^{2m}$ and $H^{\rm
odd}(X)=\prod_{m=0}^{\infty} H^{2m+1}$ are the cohomology groups in
even and odd degrees, respectively. For coefficients $\Z_q$, one has
the characteristic classes ${\overline p}_i$ as the mod q reduction
of the Pontrjagin classes $p_i$ generating the ring
$H^{**}(BO;\Z_q)=\Z_q[[{\overline p}_1, {\overline p}_2, \cdots ]]$.

\vspace{3mm} As in the case for the mod 2 classes, i.e. the
Stiefel-Whitney classes, one can form the Wu classes, and the
construction is analogous. We now have an orientation so we work
with $BSO$ rather than $BO$. By using the `inverse' $P^{-1}$ of the
operation $P$, one can define \( U(P)=P^{-1}\phi^{-1}P\phi(1) \in
H^{**}(BSO;\Z_q), \label{wu3} \) where $\phi$ is the extension to
$H^{**}(BSO;\Z_q)$ of the Thom isomorphism $\phi~:~H^{*}(BSO;\Z_q)
\longrightarrow H^{*}(MSO;\Z_q)$, and $1$ is the unit in
$H^0(X;\Z_q)$.

\vspace{3mm}
The above is indeed analogous to the
more familiar result for the mod 2 Wu class that uses the total Steenrod operation $Sq$,
\(
v(Sq)=Sq^{-1}\phi^{-1}Sq \phi(1) \in H^{**}(BO;\Z_q).
\label{wu2}
\)
Applying $Sq$ to (\ref{wu2}) gives the class $Sq v(Sq) \in
H^{**}(BO;\Z_q)$ as the
direct product of the universal Stiefel-Whitney classes.
Likewise, applying $P$ to (\ref{wu3}) gives the classes
$q_i=(P U(P))_i$ as the
direct product of the universal mod 3 classes.
The classic results of Wu imply that the classes $q_i$ are {\it
oriented} homotopy invariants and the Stiefel-Whitney classes are
homotopy invariants.

\vspace{3mm}
The Wu classes can be written
in terms of multiples of the Hirzebruch L-polynomials \cite{Hir1, Hir2}.
For every prime $q$ certain polynomials (with respect to the
cup-product) in the Pontrjagin classes $P_i$ reduced mod $q$
are topological invariants (mod $q$).
For $q=2$ of course one has the Stiefel-Whitney classes. Since
$p_i=w_{2i}^2$ (mod 2) then $p_i$ reduced mod 2 is invariant.
For $q$ an odd prime, the Steenrod powers $P_q^r$ lead to certain
polynomials $U_q^r \in H^{2r(q-1)}(M^m;\Z_q)$ in the Ponrjagin classes
which are topologically invariant, and which are characterized
by the property
\(
P_q^r (v)=U_q^r(v)~~{\rm for~~all~~}v \in H^{m-2r(q-1)}(M^m;\Z_q).
\)
As mentioned before, these can be written in terms of the Hirzebruch
L-polynomials as
\(
U_q^r=q^rL_{\frac{1}{2}r(q-1)}(p_1,p_2, \cdots)~ ({\rm mod}~~q).
\label{uqr}
\)
Thus (for $M^8$) the first Steenrod power at the prime $p=3$ is
\bea
U_3^1&=&3L_1~~~{\rm mod}~3
\nonumber\\
&=&3 \frac{p_1}{3}~~~{\rm mod}~3
\nonumber\\
&=&p_1~~~{\rm mod}~3.
\label{u31}
\eea

\section{Action of the Steenrod Reduced Powers}
\label{actionof}

Since the Steenrod reduced power operation $P_q^r$ raises the
cohomology degree by $2r(q-1)$, we see that the highest prime that
keeps us within dimensions twelve is $q=5$. The possible stable
operations in that range are \footnote{In this list we omit the
subscript $q$ as it is obvious.}

(i) {\bf $q=2$}: $Sq^i$ for $i\leq 6$,

(ii) {\bf $q=3$}: $P^1$, $\beta P^1$, $P^2$, $\beta P^2$,

(iii) {\bf $q=5$}: $P^1$, $\beta P^1$.

\noindent We are further interested only in degree four classes, that we would like
to either square or cube, and in degree seven and degree eight classes
whose degree we raise only up to a maximum of twelve.

\vspace{3mm} Let us start with the degree four class. Note that the
$\beta P_q^i$ are of odd dimension and thus are not useful in this
case. They, however are useful in type II string theory (see
\cite{ES2}) and later for the discussion of $G_7$. Thus, we are left
with only $Sq^4$ and $P_3^1$, which square a degree four class, and
with $P_3^2$ and $P_5^1$, which cube a degree four class. So we see
just from this dimensional analysis that the first pair makes up the
candidates in 8 dimensions, whereas the second two are the
candidates in 12 dimensions. Of course this analysis is only to
motivate the discussion and later we will resort to more precise
arguments that come from making the connection to Spin K-theory.

\vspace{3mm}
The Adem relation in the mod $q$ Steenrod algebra for the Steenrod powers involving $P_q^1$ is
\(
P_q^1 P_q^{2^k-1}=2^k P_q^{2^k}.
\)
Then if the dimension of the generator $x$ is $2^{k+1}$, the Adem
relation on $x$ gives
\(
x^n=\frac{1}{2^k} P_q^1 P_q^{2^k-1}x ~~({\rm mod}~q)
\)
where $x^n$ is the cup-product $n$-power of $x$, $\underbrace{x \cup x
\cdots \cup x}_{n}$. From this one can
easily get a restriction on the degree in order to have a non-zero cube.
For $q=3$,
\(
x \cup x \cup x=(-1)^k P_3^1 P_3^{2^k-1}x ~~({\rm mod}~3),
\)
and since the dimensions of $P_3^{2^k-1}x$ is $3.2^{k+1}-4$, we see that
the cube $x \cup x \cup x$ is zero (mod 3) unless $3.2^{k+1}-4$ is a
multiple of $2^{k}$. This happens only for $k=0$ and $k=1$,

{\bf (i)}{\bf $k=0$}: dim $x=2$, $x_2 \cup x_2 \cup x_2=P_3^1 x_2~~({\rm mod}~3)$

{\bf (ii)}{\bf $k=1$}: dim $x=4$, $x_4 \cup x_4 \cup x_4=\frac{1}{2}P_3^1P_3^1
x_4~~({\rm mod}~3)~ = P_3^2x_4~~({\rm mod}~3)$.

\vspace{3mm}
Let us consider the latter case, where $x_4 \in H^4(X;\Z)$ is an integral generator.
The Adem relation $P^1 P^1=2P^2$ for a general prime $q$ implies that
\(
P^1 \overline{x}_4 = \pm 2 \overline{x}_4^{\frac{(q+1)}{2}}
\label{ade}
\)
in $H^*(X;\Z_p)$, where $\overline{x}_4$ is the mod $q$ reduction of the
integral generator $x_4$. Therefore, we have

{\bf (i)}{\bf $q=3$}: $P_3^1 \overline{x}_4 = \pm 2
\overline{x}_4^2$ with $\overline{x}_4 = \rho_3(x_4)$,

{\bf (ii)}{\bf $q=5$}: $P_5^1 \overline{x}_4 = \pm 2
\overline{x}_4^3$ with $\overline{x}_4 = \rho_5(x_4)$,
where $\rho_q$, again, denotes reduction modulo $q$.

\section{ Modulo Three Reductions of the Fields}
In this section we consider the mod 3 reduction of the fields and we consider the
possible actions of the admissible cohomology operations on them.

\subsection{The Degree four field}
The first Steenrod reduced power operation for $\Z_3$ cohomology is
$P_3^1$, which takes elements in $H^{k}(X;\Z_3)$ into elements of
$H^{k+4}(X;\Z_3)$. We consider the mod 3 reduction $x_4=\rho_3(G_4)$
of the M-theory field $G_4$. We know from Ref. \cite{Flux} that
$G_4$ extends to the twelve-dimensional bounding theory on $Z^{12}$,
i.e. such that the eleven manifold $Y^{11}$ is $\partial Z^{12}$. In
this case, in addition to the first Steenrod reduced power $P_3^1$
at $p=3$ (outlined above and will be discussed further in section
(\ref{deg8})), we can also consider the second operation $P_3^2$,
which raises the cohomology degree by eight. Thus we have \( P_3^2
x_4 \in H^{12} (Z^{12}, \Z_3), \) which is equal to $U_3^2 x_4$,
where now \bea U_3^2 &=& 3^2 L_2~~~{\rm mod}~3
\nonumber\\
&=& 3^2 \frac{1}{45} (7p_2^2 - p_1^2)~~~{\rm mod}~3
\nonumber\\
&=& \frac{7p_2 - p_1^2}{5}~~~{\rm mod}~3.
\label{L2}
\eea
Thus, the action of $P_3^2$ on the mod three reduction of $G_4$ is
\(
P_3^2 x_4=\rho_3 \left( \frac{7p_2 - p_1^2}{5} \right) x_4.
\)


\vspace{3mm}
Since $G_4$ also involves a gravitational shift that involves $p_1$, we also
mention the action of power operations on the first Pontrjagin class. The mod 3
reduction of the Pontrjagin class $\rho_3(p_1(\xi))$ is an element in $H^4(X;\Z_3)$,
given by the Wu class $U_3^1(\xi)$. Thus we can have an action of $P_3^1$, and
the result is
\bea
P_3^1 \rho_3(p_1(\xi))&=&P_3^1 U_3^1(\xi)
\nonumber\\
&=& \rho_3 \left( 2p_2(\xi) - p_1^2(\xi) \right).
\label{mod3}
\eea

\subsection{The Degree seven dual field $G_7$}
We are interested in the action of cohomology operations (at $q=3$)
on the fields (reduced modulo three). Since the smallest dimension
for such an operation is four, this means that we cannot consider
the dual degree eight class $\Theta$ (in the notation of \cite{DFM})
without going beyond eleven dimensions. We can, however, consider
the differential form $G_7=*_{11}G_4$, on which we perform the mod
three reduction after lifting to an integral class. Let us call the
resulting class $x_7 \in H^7(Y^{11},\Z_3)$. In this case the first
Steenrod reduced power $P_3^1$ at $q=3$ acts on $x_7$ to give a top
class \( P_3^1 x_7 \in H^{11}(Y^{11}, \Z_3). \label{pd} \) This
top-dimensional element is characterized by the Poincar\'e duality
theorem \footnote{Note that (\ref{pd}) is a top class in $\Z_3$.
Such situations may occur (at least for homology) when the space is
not a closed manifold but rather a manifold with multiple boundary
components together with an identification of these components. A
standard class of examples is the so-called $\Z_k$- (or $\Z/k$-)
manifolds of Sullivan.}
 and is given
by the class $U_3^1 x_7$. The element $U_q^r$ is given by (\ref{uqr}).
Adapting to our situation, with $p=3$, $r=1$, we have
the Wu class $U_3^1$ (eqn. (\ref{u31})). Therefore, the action of $P_3^1$ on the mod 3 reduction of $G_7$ is given by
\(
P_3^1 x_7=\rho_3(p_1) \cup x_7=U_3^1\cup x_7.
\label{x7}
\)

\subsection{The Degree eight `dual' field $\Theta$}
\label{deg8}
Here we would like to act by cohomology operations on the mod 3 reduction $\rho_3(\Theta)=y_8$
of the class $\Theta$. Assuming the class extends to twelve dimensions, we can consider
\(
P_3^1 y_8 \in H^{12}(Z^{12}, \Z_3).
\)
As in the case for $G_7$  this is equal to $U_3^1 y_8$, so that
\(
P_3^1 \rho_3(\Theta) = \rho_3(p_1) \rho_3(\Theta),
\)
which is analogous to (\ref{x7}).

\vspace{3mm}
Next we show that the degree eight class $\Theta(\rho_3(a))$ corresponding to
the mod 3 reduction can be written as a cohomology
operation. We use (\ref{ade}) and the additivity of the mod $k$ reduction,
i.e. $\rho_k(a+b)=\rho_k(a) + \rho_k(b)$,
to calculate for $G_4$ reduced mod 3, $\overline{G_4}$, the following\footnote{Here we assume that $\overline{G}_4$ is in cohomology. This would come
from assuming both factors in the shifted quantization condition \cite{Flux} to be in
integral cohomology, an so the mod $q$ reduction is in mod $q$ cohomology.}
\bea
\frac{1}{2} \left[ \frac{1}{2}P_3^1 {\overline G_4} + {\overline G_4} \cup
{\overline G_4}
\right]
&=&
\frac{1}{2} \left[ \frac{1}{2}P_3^1 \left( {\rho}_3(a) - {\rho}_3(\lambda/2) \right)
+
\left( {\rho}_3(a) - {\rho}_3(\lambda/2) \right) \cup \left( {\rho}_3(a) -
{\rho}_3(\lambda/2) \right)
\right]
\nonumber\\
&=& {\rho}_3(a) \cup {\rho}_3(a) - {\rho}_3(\lambda/2)
\cup {\rho}_3(\lambda/2)
+ {\rho}_3(a) \cup {\rho}_3(a)
\nonumber\\
&&+ {\rho}_3(\lambda/2) \cup {\rho}_3(\lambda/2)
- {\rho}_3(a) \cup {\rho}_3(\lambda)
\nonumber\\
&=& \left[ 2 {\rho}_3(a) \cup {\rho}_3(a) - {\rho}_3(a) \cup {\rho}_3(\lambda)) \right]
\nonumber\\
&=& \Theta\left( {\rho}_3(a) \right), \label{main} \eea the DFM
class with the degree eight term set to zero. The full result will
involve the the reduction of $I_8$. The division by two on he left
hand side is harmless since we are reducing modulo 3. This may be
thought of as mod 3 analog in M-theory of the mod 2 expression in
type II string theory, namely the Freed-Witten anomaly cancelation
formula for D-branes \cite{FW} $(Sq^3 + H_3\cup)F=0$, since the
class $\Theta$ measures the anomaly of the M-branes \cite{DFM}.

\section{The Quadratic Refinement and Spin K-theory}
\label{KSp}

In this section we will show that the multiplicative structure on
the degree four and degree eight cohomology encodes the quadratic
refinement law of \cite{DFM} for the eight-form in M-theory, the
refinement being given by the cup product of two 4-forms from $G_4$.
We will see that this is reflected in the addition on the target
(equation (\ref{quad})).

\vspace{3mm}
The degree eight class in M-theory is given by the integral lift
of the (negative of the) right hand side of the equation of motion
for $G_4$, which is
\(
d*G_4=-\frac{1}{2}G_4 \wedge G_4 + I_8,
\)
so that the degree eight class $\Theta(a)$, defined in \cite{DFM}, is
\(
\Theta(a)=\left[ \frac{1}{2}G_4 \wedge G_4 - I_8 \right],
\)
whose expression in terms of integral classes $a$ and $\lambda$ reads
\(
\Theta(a)=\frac{1}{2}a(a-\lambda) + 30 {\widehat{A}}_8.
\)

\vspace{3mm}
Among the properties of this class proved in \cite{DFM} is that it is
a quadratic refinement of the cup product of two degree four classes
$a_1$ and $a_2$
\(
\Theta(a_1+a_2) + \Theta(0)= \Theta(a_1)  + \Theta(a_2)+ a_1 \cup a_2.
\label{ref}
\)
We would like to look at this from the point of view of the structure on
the product of the cohomology groups $H^4(~;\Z) \times H^8(~;\Z)$.
For this we consider the two classes $a$ and $\Theta(a)$ as a pair
$(a,\Theta(a))$ in $H^4(~;\Z) \times H^8(~;\Z)$. Then the linearity
of the addition of the degree four classes $a$ and the quadratic
refinement property (\ref{ref}) of $\Theta(a)$ can both be written in
one expression in the product $H^4(~;\Z) \times H^8(~;\Z)$, which
makes use of the ring structure, namely
\(
\left( a_1,\Theta(a_1) \right) + \left( a_2,\Theta(a_2) \right)=
\left( a_1 + a_2, \Theta(a_1) + \Theta(a_2) + a_1 \cup a_2 \right).
\label{t2}
\)
The second entry on the RHS is just $\Theta(a_1+a_2) - \Theta(0)$,
and so it encodes the property (\ref{ref}).

\vspace{3mm}
We can define the shifted class $\Theta^0(a)$ as the difference
$\Theta(a) - \Theta(0)$, so that (\ref{t2}) is
replaced by
\(
\left( a_1,\Theta^0(a_1) \right) + \left( a_2,\Theta^0(a_2) \right) =
\left( a_1 + a_2, \Theta^0(a_1+a_2)\right),
\label{t3}
\)
corresponding to the special case
\(
\Theta^0(a_1+a_2) = \Theta^0(a_1)  + \Theta^0(a_2)+ a_1 \cup a_2.
\)
This is then just a realization of the multiplication law on
$H^4(~;\Z) \times H^8(~;\Z)$ which, for $(a,b)$ in the product group,
is 
\(
(a_1, b_1) + (a_2, b_2)=(a_1 + a_2, b_1 + b_2 + a_1 \cup a_2).
\label{quad}
\)
Note that in order to get this law we had to use the modified
eight-class $\Theta^0(a)$, or alternatively discard $\Theta (0)=30
{\widehat{A}}_8$.
\footnote{One way is to set this to zero rationally by requiring
$p_2$ to be equal to $\frac{7}{4}p_1^2$, but this does not seem to
be the best possible.}
From the quadratic refinement law, \cite{DFM} noted that this term can
at most be two-torsion.

\vspace{3mm}
We now make the connection to Spin K-theory.
Similarly to the case of other kinds of bundles, e.g. complex or real,
one can get a Grothendieck group of isomorphism classes of Spin
bundles up to
equivalence. The reduced $K{\rm Spin}$ group of a
topological space can be defined as
$\widetilde{K{\rm Spin}}(X)=[X,B{\rm Spin}]$.
For the case of $B{\rm Spin}$, we will be interested in relating Spin
K-theory to cohomology of degrees 4 and 8. Such a homomorphism of
abelian groups
\( Q_X:{\widetilde{K{\rm Spin}}}(X) \rightarrow H^4(X;\ZZ) \times
H^8(X;\ZZ)
\label{ch}
\)
is defined by \cite{Li}
$Q_X\left( Q_1(\xi),Q_2(\xi)\right)$ for $\xi \in
{\widetilde{K{\rm Spin}}}(X)$. We see that this is the Spin analog of (\ref{KO}).
For two bundles $\xi$ and $\gamma$ in
${\widetilde{K{\rm Spin}}}(X)$, and for $k\leq 3$,
\( Q_k(\xi \oplus \gamma)=\sum_{i+j=k}Q_i(\xi) \cup Q_j(\gamma).
\label{add}
\)
We also see that the map (\ref{ch}) is essentially our `gravitational' Chern character in
\cite{S1}. The fact that this relation only works for
$k\leq 3$ is in accord with the observation that the expressions in
\cite{S1} also only work for that range.
\footnote{We thank Michael Hopkins and Isadore Singer for pointing out to
us that from a topological point of view, such a structure also only works
in low degrees, interestingly in the range of dimension relevant to
M-theory.}
The addition on the target is given precisely by (\ref{quad})
for $(a,b)\in H^4(X;\ZZ) \times H^8(X;\ZZ)$ \cite{Thomas} .

\vspace{3mm}
The quantization condition on $G_4$ \cite{Flux} (see the introduction) involves
an integral class coming from the $E_8$ bundle. How does this $E_8$ part
fit into the above discussion? Since $H^8(E_8)=0$, then any degree eight class would have
to come from the only class of lower degree, namely the degree four class. The only
possibility is squaring. Indeed, using Chern-Weil representatives,
${\rm Tr}F^4=\frac{1}{100}{\rm Tr}\left( F^2 \right)^2$. This implies that that the only degree eight class comes in the form of a composite,
$a_1 \cup a_2$ for $a_1$ and $a_2$, the generators of $H^4(X,\Z)$ pulled back
from  $H^4(BE_8, \Z)$.

\section{Realizing the Anomalies in this Approach}
Given an action $S$ in Euclidean signature, it often splits into a
real and an imaginary parts, $S= {\rm Re} S + i {\rm Im} S$, so that
when forming the semi-classical partition function
$\int_{\mathcal{M}} e^{2\pi i S}$ one gets a modulus and a phase.
The latter is usually given by the topological (i.e. the
metric-independent) parts $S_{\rm top}$ of the action as ${\rm
Phase}~= e^{2\pi i {\rm Re} S}=e^{2 \pi i S_{top}}$. In studying the
topological aspects of the partition function in M-theory, and upon
including torsion fields, this phase leads to subtle signs that give
potential anomalies. In \cite{DMW} the condition on the phase ended
up being that it is essentially identically one. That involved the
study of the divisibility properties of the fields. Since this lives
in $\Z_2$, the phase was just given by the mod 2 reduction of the
action, which by Witten's earlier result \cite{Flux} is just the sum
of the mod 2 index of the Dirac operator coupled to an $E_8$ bundle
and the mod index of the Rarita-Schwinger operator, i.e. the Dirac
operator coupled to the tangent bundle (minus 3 copies of the
trivial line bundle). Explicitly \cite{Flux} \cite{DMW} \( \Phi=\exp
2\pi i \left[ \frac{1}{2} {\rm Index}(D_{E_8}) + \frac{1}{4} {\rm
Index} (D_{R.S.}) \right]. \) Using the Atiyah-Patodi-Singer index
theorem and using the fact that the mod 2 index of the Dirac
operator coupled to a real bundle in ten dimensions is a topological
invariant, the phase was shown by Witten to reduce to $\Phi =
(-1)^{f(a)}$, where $f(a)$ is the mod 2 index of the Dirac operator
coupled to the $E_8$ vector bundle with a degree four class $a$. In
\cite{DMW} this mod 2 index was studied via torsion pairings on
cohomology. On $X^{10}$ and two degree four classes $a,b \in
H^4(X^{10};\Z)$, the torsion pairing used is $T(a,
Sq^3b)=\int_{X^{10}} a \cup Sq^2 b$,  where by Adem relation, $\beta
(Sq^2 b)= Sq^3 b$. In general $T$ takes values in $U(1)$ but in this
case it takes values in $\Z_2 \subset U(1)$ since $Sq^3b$ is
2-torsion. The mod 2 index $f(a)$ is a quadratic refinement of the
bilinear form via the cup product \cite{DMW} \( f(a_1 + a_2) =
f(a_1) + f(a_2) + \int_{X^{10}} a_1 \cup Sq^2 a_2. \label{mod2index}
\)

\vspace{3mm} First, note that we have written $I_8$ in terms of the
Spin characteristic classes. In particular, the expression
(\ref{Q24}) for $I_8$ includes $Q_2$, so in order to look at a
possible mod 2 reduction of $I_8$ we need to see what the
corresponding reduction of the $Q_i$'s is. The mod 2 reduction $r_2$
of the Spin classes are the Stiefel-Whitney classes in that
dimension, i.e. \bea
\rho_2(Q_1)&=& w_4 \nonumber\\
\rho_2(Q_2)&=& w_8. 
\label{w4w8}
\eea However, we see that we have the division
by $24$ which makes the task nontrivial. \footnote{One might be able
to evade this subtlety by looking at the integral of the one-loop term
(\ref{I8}) lifted as usual to a twelve-dimensional bounding Spin
manifold $Z^{12}$. If we assume that the class of $G_4$ is divisible
by 24 then we can write that integral as  $\int_{Z^{12}}
\frac{G_4}{24}\wedge Q_2$, assuming $G_4$ is in cohomology.}

\vspace{3mm} The presence of the one-loop term in M-theory $
\int_{Y^{11}} C_3\wedge I_8$ reduced in type IIA
string theory to the corresponding one-loop term $\int_{X^{10}}
B_2\wedge I_8$. Similarly, the Chern-Simons term
$\frac{1}{6} \int_{Y^{11}} C_3 \wedge G_4 \wedge G_4$ reduces to
the corresponding Chern-Simons term in type IIA $ \frac{1}{6}
\int_{X^{10}} B_2 \wedge F_4 \wedge F_4$.
The field $F_4$ is obtained from the M-theory field
$G_4$ and so is expected to also have a shift proportional to
$Q_1$, the mod 2 reduction of which is $w_4$. Now the mod 2
reduction of the action amounts to replacing the fields by their mod
2 reductions, together with the mod 2 Steenrod operations
\footnote{We could have included $w_2$ with $B_2$, but we are
assuming our ten-manifold to be spin.}, so schematically
 $F_4$ should correspond to $w_4$ and $Sq^4$, $I_8$ to $w_8$,
 and $B_2$ to $Sq^2$. Now we take $B_2$ to correspond to a cohomology
 operation given by the second
Steenrod Square $Sq^2$ (that is how it shows up in KO-theory), and
so the operation replacing the one-loop term is $\int_{X^{10}} Sq^2
I_8$. By using (\ref{Q24}) we see that the condition is
\footnote{This involves mod 2 reduction implicitly.} $Sq^2 Q_2=0$.
Thus form the topological action we get three possible terms in the
mod 2 reduction, namely $w_4 Sq^2 w_4$, $Sq^2 Sq^4 w_4$, and $sq^2
w_8$. In what follows we will show that such terms correspond
naturally to expressions in Spin K-theory (see (\ref{q2})). The
dimensions relevant here are: four for the M2-brane theory, eight
for the M5-brane theory, ten for type II string theory, and twelve
for M-theory (more precisely, the cobounding theory).

\subsection{The Fivebrane and eight-manifolds}

The topological part of the M5-brane action extended via the
Chern-Simons construction from six dimensions to eight dimensions is
given by \cite{5} \cite{HS} \( S_8 = \frac{1}{2} \int_{M^8} G_4
\wedge G_4 - \lambda \wedge G_4. \) The mod two reduction of this
action is \( {\rho}_2(a) \cup {\rho}_2(a) - v_4 \cup {\rho}_2(a), \)
where we denote by $\rho_2(a)$ the mod two reduction of the integral
class $a$ of $G_4$, and $U_2^1$ is the second Wu class \footnote{In
this general notation, $U_2^1$ corresponds to $v_1$ or $\nu_1$ in
the notation more particular to the prime 2.} given in terms of the
Stiefel-Whitney classes by the Wu formula $U_2^1=w_4 - w_2^2$. For
$M^8$ spin, which is what we assume, then $U_2^1$ is the same as
$w_4$. Similarly, the mod three reduction takes the form \(
\rho_3(a) \cup \rho_3(a) - U_3^1 \cup \rho_3(a), \) where
$\rho_3(a)$ denotes the mod three reduction of the integral class
$a$, and $U_3^1$ is the first Wu class at the prime $p=3$.

\vspace{3mm} Consider the exact sequence \cite{Li} \( 0
\longrightarrow {\rm ker}Q_1 \longrightarrow
{\widetilde{K{\rm Spin}}}(M^8) {\buildrel{Q_1} \over {\longrightarrow}}
H^4(M^8;\Z) \longrightarrow 0. \label{ker} \) Since the kernel of
$Q_1$ is string manifolds, then we see that the difference between
this Spin K-theory and integral four cohomology is the string
condition. The Spin K-theory picks degree four classes that are in
the image of $Q_1$ modulo the ones in its kernel. Since this looks
like cohomology then it makes sense to expect to be able to replace
$Q_1$ by some cohomology operation that would appear in the
corresponding Atiyah-Hirzebruch spectral sequence. We further ask
the question: what is the meaning of $Q_2$ once $Q_1$ vanishes, i.e.
for String manifolds? The existence of the exact sequence, which is
an isomorphism, \cite{Li} \( Q_2|_{{}_{{\rm ker}Q_1}}~:~{\rm ker}Q_1
\longrightarrow 3H^8(M^8;\Z) \label{3} \) means that once $Q_1$ is
zero, $Q_2$ coincides with three times the eighth integral
cohomology of the manifold. Since in this case $Q_2$ would be just
twice the second Pontrjagin class, $2p_2$, then this implies that
$p_2$ is equal to six times the integral generator. Thus we see that
for a String manifold, the second Pontrjagin class is divisible by
six. This is obviously consistent with the divisibility by two in
the proposal in \cite{S1}.

\subsection{The Mod 2 anomaly}

The discussion leading to the mod 2 reduction of the action involved
only the $E_8$ classes and did not include the gravitational class
$\lambda/2$ appearing in the shifted quantization condition for the
M-theory four-form (\ref{g4}). In particular, they involved the Wu
relations among the Chern classes of the unitary bundle obtained
from the breaking $E_8 \supset \left( SU(5) \times
SU(5)\right)/\Z_5$ \cite{DMW}. In our present context of Spin
characteristic classes, we would like to give the corresponding
condition on these classes. Since $\lambda/2$ appears linearly with
$a$, the Spin classes will have an analogous expression
\footnote{Again the mod 2 reduction is understood implicitly.} $Q_1
\cup Sq^2 Q_1$. We would like to investigate whether this can be
obtained in a systematic way as part of an expression in KSpin which
would also have a topological interpretation. In a given dimension,
there are relations between the characteristic classes and the
cohomology operations. In this case, the relations in
$H^{10}(BSO;\Z_2)$ are given as linear combinations of the possible
Steenrod square operations acting on the generators (\ref{w4w8}), namely
$w_2^2\cup Sq^2w_2^2$, $Sq^4Sq^2w_2^2$, and $Sq^2w_4^2$. In the spin
case, only the latter survives.

\vspace{3mm} We are dealing with degree four and degree eight class
so we can pull back the above classes to the classifying spaces
$K(\Z,4)$ and $K(\Z,8)$, since cohomology groups of $X$ can be
understood as the homotopy classes of maps from that space to the
Eilenberg-Maclane spaces \bea H^4(X,\Z) \times H^8(X,\Z)&=&\left[
X,K(\Z,4)\right] \times \left[ X,K(\Z,8)\right]
\nonumber\\
&=&\left[ X,K(\Z,4)\times K(\Z,8) \right]. \label{KZ48} \eea Let
$x\in H^4(K(\Z,4),\Z)$ and $y\in H^8(K(\Z,8),\Z)$ be the standard
generators. We are further interested in classes in $\Z_2$, so let
the corresponding mod 2 reductions be given by \bea z_4=x~{\rm
mod}~2 \in H^4(K(\Z,4),\Z_2)
\nonumber\\
z_8=y~{\rm mod}~2 \in H^8(K(\Z,8),\Z_2). \eea In the Postnikov tower
with lowest level $E_0$, $H^{10}(E_0;\Z_2)$ as a vector space over
$\Z_2$ has a basis $z_4\cup Sq^2 z_4$, $Sq^4Sq^2z_4$, and $Sq^8z_8$.
It turns out that the coefficients in the linear combination are one
so that the second $k$-invariant is given by \cite{Li} $k_2=z_4\cup
Sq^2 z_4+ Sq^4Sq^2z_4 + Sq^8z_8$, and the corresponding map \(
\Lambda_X: H^4(X;\ZZ) \times H^{8}(X;\ZZ) \rightarrow
H^{10}(X;\ZZ_2), \) is given by \cite{Li} \( \Lambda_X(Q_1,Q_2)=Q_1
\cup Sq^2 Q_1 + Sq^4 Sq^2 Q_1 + Sq^2 Q_2. \label{q2} \) We view this
map as the mod 2 index for Dirac operators coupled to Spin bundles,
and the vanishing of the mod index is then essentially
\footnote{Note that there are factors of half involved.} the
condition to lift the degree four (and eight) cohomology to Spin
K-theory.

\subsection{The DFM anomaly}
In this section we look at the DFM anomaly \cite{DFM}. We aim at
achieving two things: First, encode the structure of the degree four
and degree eight classes in our context of Spin characteristic
classes, and second, seek at a possible variant of this anomaly to
include mod three reductions of fields. The first was considered in
(\ref{deg8}), so here we consider the second.

\vspace{3mm} In order to describe the electric charge induced by the
self-interactions of the C-field, Ref. \cite{DFM} defined an
integral lift of the EOM of $G_4$, $\Theta_X(a)$, where $a$ is the
integral class appearing in the shifted quantization condition of
$G_4$ (\ref{g4}). We note that the quadratic refinement is exactly
the addition law on the target of the map $Q_X$ in (\ref{quad}).
Thus we see that the product of the two cohomology groups $H^4$ and
$H^8$ together with their ring structure encodes the elements $a$
and $\Theta_X(a)$ together with the correct addition laws. Now that
we have seen that we have the correct structure for the elements and
their addition law, we would like to see what consequence that has
on the anomaly itself.

\vspace{3mm} Let us first motivate the problem heuristically from
the point of view of ten-dimensional type IIA. There, the
Freed-Witten anomaly reads \cite{FW} $Sq^3 F + H_3 \cup F=0$, where
$F$ is the total Ramond-Ramond field strength that includes the
fields of all even degrees. Since the `operator' $Sq^3+ H_3\cup$
appearing in this equation is of a uniform degree, we can isolate
one of the RR fields. We thus focus on $F_4$, in which case $Sq^3
F_4 + H_3 \cup F_4=0$. We use this expression to get hints about
what a possible `$S^1$-lift' might be in M-theory. Since the
diagonal lift of $H_3$ as well as the vertical lift of $F_4$ to
M-theory both give $G_4$, a candidate expression in M-theory would
involve replacing $F_4$ and $H_3$ both with $G_4$, i.e.
schematically \( {\mathcal{O}}G_4 + G_4 \cup G_4, \) where
$\mathcal{O}$ is a cohomology operation we have been arguing for its
existence and which need to be determined. Again, in order to get an
equation of homogeneous degree -- that is the only choice that seems
to be available-- the operation $\mathcal{O}$ should be of degree
four, i.e. it should raise the cohomology degree by four. What are
the candidates? It seems to be only $Sq^4$ (and decomposables) at
$p=2$ or $P_3^1$ at $p=3$.

\vspace{3mm} We would like to understand the cohomology groups
$H^{4i}(X,\Z)$ for $i=1,2$ in order to understand the map from Spin
K-theory and the corresponding obstructions to lifting. We follow
\cite{Li} for the mathematical results for what follows. Given the
universal Spin characteristic classes $Q_i \in H^{4i}(B{\rm
Spin};\Z)=\left[ B{\rm Spin},K(\Z,4i) \right]$, we can pull them
back to the space $X$. To understand the image of $Q_X$ we ask which
map $f : X \longrightarrow K(\Z,4)\times K(\Z,8)$ admits a lifting
relative to the pair $Q=(Q_1,Q_2)$, \( B{\rm Spin}
{\buildrel{\Delta} \over {\longrightarrow}} B{\rm Spin} \times B{\rm
Spin} {\buildrel{Q_1 \times Q_2} \over {\longrightarrow}}
K(\Z,4)\times K(\Z,8). \) It is here that the Steenrod power
operations $P_3^1$, taking $H^4(K(\Z,4); \Z_3)$ to $H^8(K(\Z,4);
\Z_3)$, make their appearance as follows. Let $x_4$ and $y_8$ be the
standard generators of $H^4(K(\Z,4);\Z)$ and $H^8(K(\Z,8);\Z)$,
respectively. Then as vector spaces over $\Z_3$, $H^8(K(\Z,8);\Z_3)$
is generated by a single element $y_8 ~{\rm mod}~ 3$, while
$H^8(K(\Z,4);\Z_3)$ is generated by the two elements \bea
&&x_4^2~{\rm mod}~3 ~~~~~~~~~~~{\rm (decomposable)}
\nonumber\\
&&P^1(x_4~{\rm mod}~3) ~~~~~~~~~{\rm (primitive)} \eea The invariant
$k_1$ is a cohomology class that lies in \( H^8(K(\Z,4)\times
K(\Z,8);\Z_3)=H^8(K(\Z,4); \Z_3) \oplus H^8(K(\Z,8); \Z_3), \) and
so its expression is given as a linear combination of the above
three $\Z_3$-valued generators. It turns out again that the
coefficients are all one so that the map \( R_X: H^4(X;\ZZ) \times
H^8(X;\ZZ) \rightarrow H^8(X;\ZZ_3) \) given by \( R_X(a,b)=(a\cup a
+ b)~ {\rm mod}~3 + P_3^1(a~ {\rm mod}3),\) is a homomorphism, with
the group structure being that on $H^4(X;\ZZ) \times H^8(X;\ZZ)$.
\footnote{In going from $k_1$ to $R_X$ we replaced $x_4$ by $a$ and
$y_8$ by $b$.}

\vspace{3mm} The lifting condition in dimension eight is the
following \cite{Li}. The stable classes of Spin bundles over an
eight-dimensional closed manifold are in one-to-one correspondence
with pairs $(a,b)\in H^4(X;\ZZ) \times H^8(X;\ZZ)$ satisfying \(
(a\cup a + b)~{\rm mod}~3 + U_3^1 \cup (a~ {\rm mod}~3)=0, \) where
$U_3^1$ is the corresponding Wu class. It is this formula that we
think of as the mod 3 analog of the DFM formula.


\section{Further Remarks}

\noindent {\bf{The integral anomaly:}}

\vspace{1mm}
For a torsion class $c$, $f(a+2c)=f(a)+ \int c \cup Sq^2 \lambda$
\cite{DMW}. The absence of the refinement implies that in the
torsion pairing between a 4-class and a seven-class that $\beta Sq^2
\lambda$ be equal to zero. This is $T(b, Sq^3 \lambda)=0$ giving the
$W_7$ anomaly canceled in \cite{KS1} via elliptic cohomology. The
cohomology ring of $B{\rm Spin}$ over the integers contains, in
addition to the Spin characteristic classes $Q_1$ and $Q_2$ of
dimensions 4 and 8 respectively, a characteristic class of degree
seven. This is the generator of \( H^7(B{\rm Spin};\Z)=\Z_2,
\label{7} \) which is nothing but the Seventh integral
Stiefel-Whitney class $W_7$, obtained as the Bockstein on the sixth
mod 2 Stiefel-Whitney class $w_6$. This is precisely the anomaly
that DMW found \cite{DMW}. It was canceled in \cite{KS1} by
declaring the spacetime to be orientable with respect to Landweber's
elliptic cohomology $E(2)$ or Morava K-theory $K(2)$ (both taken at
the prime $p=2$), a result which was obtained by identifying $W_7$
as the cohomology class corresponding to an obstruction, i.e. as a
differential in the Atiyah-Hirzebruch spectral sequence. From
(\ref{7}) it seems that there is another interpretation, namely that
the vanishing of $W_7$ is simply the vanishing of the seventh Spin
characteristic class pulled back to spacetime from the universal
bundle $B{\rm Spin}$. Thus, the DMW anomaly can also be naturally
interpreted in this context.

\vspace{3mm}
\noindent {\bf{The $w_4$ anomaly:}}

\vspace{1mm}
This anomaly was physically proposed and mathematically derived in
\cite{KS1}. This also shows up in an apparently different context, namely
as part of the shift in the quantization of the M-theory fieldstrength \cite{Flux}.
We make a connection between the two.
We start with the following observation. If $w_4=0$ then the first Spin
characteristic class is divisible by two.
Since $Q_1 \equiv w_4$ mod $2$, then $w_4=0$ implies that $Q_1 \equiv 0$
mod 2, which implies that $Q_1$ is divisible by two. So there is some
(not necessarily unique) class $\gamma$ such that $2\gamma=Q_1$. This
gives an interpretation of the $EO(2)$ condition as giving the shift in
Witten's quantization (\ref{g4}) to be even. In this case, the
membrane path integral can be defined with no ambiguity. Thus, the $w_4$
condition, when traced back, can be viewed as the condition for an anomaly free membrane
partition function. In \cite{KS1} this was needed to construct the mod 2 part of
the generalized cohomology partition function. Thus, we interpret the construction
in \cite{KS1} as corresponding to the case when the M-theory fieldstrength
satisfied a direct quantization condition, i.e. one that is not shifted.
Note that $W_7$ is obtained from $w_4$ via the Steenrod operation
$Sq^3$. By the Wu formula $w_6=Sq^2 w_4 + w_2 w_4$, so that for spin
bundles one has $W_7=\beta Sq^2 w_4= Sq^3 w_4$, where $\beta$ is the
Bockstein map.

\vspace{3mm}
\noindent {\bf{Mod 4 reduction:}}

\vspace{1mm}
The inclusion $i : \Z_2 \to \Z_4$ induces the mapping
$i_* : H^*(X;\Z_2) \to H^*(X;\Z_4)$. For a vector bundle $\xi$, the
reduction mod 4 of the Pontrjagin classes $p_i(\xi)$ can be written
in terms of the Stiefel-Whitney classes $w_i(\xi)$ (of various
degrees) by using $\i_*$ above and the Pontjagin square
$\mathcal{P}$. This latter is a cohomology operation from
$H^{2k}(X;\Z_2)$ into $H^{4k}(X;\Z_4)$. The mod 4 reduction of the
Pontrjagin classes is \bea \rho_4 p_1(\xi)&=& {\mathcal{P}} w_2(\xi)
+ i_* w_4(\xi),
\nonumber\\
\rho_4p_2(\xi)&=& {\mathcal{P}}w_4(\xi) + i_*\left\{  w_8(\xi) +
w_2(\xi)  w_6(\xi) \right\}. \label{mod4} \eea Thus the mod 4
reductions are given essentially by the mod 2 reductions. Note that
for a Spin bundle, $w_2(\xi)$ is zero, and requiring further the
$EO(2)$ orientation condition $w_4=0$ \cite{KS1, KS3} then implies
that the mod 4 reduction of $p_1$ is zero. This would also be true
for $p_2$ if in addition we require $w_8$ to be zero, i.e. that the
second Spin characteristic class $Q_2$ used earlier is even.

\vspace{3mm}
\noindent {\bf{Mod 5 reduction:}}

\vspace{1mm}
From the definition of the Steenrod reduced powers we see that the
operation $P_5^1$ cubes a degree four class. Thus, on the mod 5
reduction $\rho_5(G_4)$ we have $P_5^1\left(  \rho_5(G_4) \right)=
\rho_5(G_4)\cup  \rho_5(G_4) \cup \rho_5(G_4)$, thus generating the
form of the cubic Chern-Simons term. What about the reduction of
$I_8$ mod 5? If we assume for simplicity that $p_1/2=0$, then $I_8$
reduces to $p_2/48$, the mod 5 reduction of which we write as
$p_2/2$ mod $120$. The Pontrjagin classes mod 120 are topological
invariant \cite{Singh91}. If we use Spin bundles and their higher
connected analogs then the right classes to look at are the Spin
characteristic classes formed of $p_1/2$ and $p_2/2$. We expect that
using these classes we get the topological invariance of $I_8$
reduced modulo 5.

\vspace{3mm}
\noindent {\bf{Type II and the AHSS:}}

\vspace{1mm}
In type IIA string theory it was argued in \cite{ES2} that a D-brane which is
free of Freed-Witten anomalies lifts to twisted K-theory if and only if the
Poincar\'e dual of the cycle that it wraps is annihilated by the Milnor primitive
$Q_1=-\beta P_3^1$. This operator is indeed the fifth
differential $d_5$ in the Atiyah-Hirzebruch Spectral Sequence for complex
K-theory at $q=3$. Mathematically, this follows from \cite{Bu} where the differentials
at prime $q \geq 2$ are given by $d_{2r(q-1)+1}=\beta P_q^r$. For $q=3$
we see that the first differential is just the Bockstein $\beta$ and the third
is $d_9=\beta P_3^2$.  This shows that the only nontrivial operation at
$q=3$ in string theory is $\beta P_3^1$ considered in \cite{ES2}. In light of this
discussion, there does not seem to be anything special about $q=3$
in the considerations in \cite{ES2} except providing examples and
staying within the allowed range of dimension. This suggests that
$q=5$ examples should be relevant in type II but they have to be
restricted to degree one classes, as seen by the fact that $P_5^1$
raises the cohomology degree by eight.

\vspace{3mm} In our current M-theory context, the formula (\ref{main}) 
suggests an obstruction in a spectral sequence for which we argued earlier.  
The differential has order four. Even differentials
are usually associated with real (rather than complex) theories--
for example, whereas the first differential for K-theory is
$d_3=\beta Sq^2$, for $KO$-theory it is $d_2=Sq^2$, and this
generalizes to other theories as well-- and so this is compatible with 
the requirement that the theory be real.

\vspace{3mm} In closing we point out that a more careful account for denominator
factors is needed. We have not been precise on those. However, we expect, in 
line of previous work, that accounting for factors such as 24 will make contact with
higher $BO\langle n \rangle$. This will be the subject of the next
step in our investigation.

\bigskip\bigskip
\noindent
{\bf \large Acknowledgements}\\
\vspace{2mm}
\noindent
We thank Matthew Ando for useful discussions.


\end{document}